\documentclass[traditabstract]{aa} 

\usepackage{graphicx,longtable,cancel}
\usepackage{txfonts}
\usepackage{natbib}
\usepackage{rotating}
\usepackage{longtable}
\usepackage{supertabular}
\usepackage{float,lscape}
\begin{document}
\bibliographystyle{aa}

\makeatletter
\def\@biblabel#1{\hspace*{-\labelsep}}
\makeatother

\title{The fine line between total and partial tidal disruption events}

\author{
Deborah Mainetti$\rm ^{1, 2, 3}$, Alessandro Lupi$\rm ^{3, 4}$, Sergio Campana$\rm ^{2}$, Monica Colpi$\rm ^{1, 3}$, Eric R. Coughlin$\rm ^{5, 6}$, James Guillochon$\rm ^{7}$ and Enrico Ramirez-Ruiz$\rm ^{8}$ 
}
\institute{
Dipartimento di Fisica G. Occhialini, Universit\`a degli Studi di Milano Bicocca, Piazza della Scienza 3, I-20126 Milano, Italy
\and
INAF, Osservatorio Astronomico di Brera, Via E. Bianchi 46, I-23807 Merate (LC), Italy
\and
INFN, Sezione di Milano-Bicocca, Piazza della Scienza 3, I-20126 Milano, Italy
\and
Institut d'Astrophysique de Paris, Sorbonne Universit\`{e}s, UPMC Univ Paris 6 et CNRS, UMR 7095, 98 bis bd Arago, F-75014 Paris, France
\and
Astronomy Department and Theoretical Astrophysics Center, University of California, Berkeley, Berkeley, CA 94720
\and
Einstein Fellow
\and
Harvard-Smithsonian Center for Astrophysics, The Institute for Theory and Computation, 60 Garden Street, Cambridge, MA-02138, USA
\and
Department of Astronomy \& Astrophysics, University of California, Santa Cruz, CA-95064, USA
              }
              \date{\email{d.mainetti1@campus.unimib.it}}
              
              \abstract{Flares from tidal disruption events are unique tracers of quiescent black holes at the centre of galaxies. The appearance of these flares is very sensitive to whether the star is totally or partially disrupted, and in this paper we seek to identify the critical distance of the star from the black hole ($r_{\rm d}$) that enables us to distinguish between these two outcomes. We perform here mesh-free finite mass, traditional, and modern smoothed particle hydrodynamical simulations of star-black hole close encounters, with the aim of checking if the value of $r_{\rm d}$ depends on the simulation technique. We find that the critical distance (or the so-called critical disruption parameter $\beta_{\rm d}$) depends only weakly on the adopted simulation method, being $\beta_{\rm d}=0.92\pm 0.02$ for a $\gamma=5/3$ polytrope and $\beta_{\rm d}=2.01\pm 0.01$ for a $\gamma=4/3$ polytrope.                 
              }

              \keywords{hydrodynamics - methods: numerical - galaxies: nuclei - black hole physics - accretion, accretion disks}
               \authorrunning{Mainetti et al.} 
\maketitle

\section{Introduction}
There is compelling evidence of the ubiquitous presence of massive black holes (BHs) at the centres of nearby galaxies (Kormendy \& Richstone 1995; Kormendy \& Ho 2013). 
These BHs are mainly in low-luminous states (Ho 2008) or in quiescence, but sometimes they can enter highly luminous phases (AGN) that are due to sudden influxes of the surrounding gas. These influxes can be provided by the tidal disruption (TD) of stars (Rees 1988). TDs occur when stellar dynamical encounters scatter a star (of mass $M_{\rm *}$ and radius $R_{\rm *}$) onto a low angular momentum orbit about the BH (of mass $M_{\rm BH}$), subjecting it to the extreme BH tidal field (Alexander 2012). Specifically, if the star comes close to the so-called BH tidal radius
\begin{equation}
r_{\rm t}\sim R_{\rm *}\Biggl(\frac{M_{\rm BH}}{M_{\rm *}} \Biggr)^{1/3}\sim 10^2 \rm R_{\rm \odot} \Biggl(\frac{\it R_{\rm *}}{1 \rm R_{\rm \odot}}\Biggr)\Biggl(\frac{\it M_{\rm BH}}{10^6 \rm M_{\rm \odot}}\Biggr)^{1/3}\Biggl(\frac{1 \rm M_{\rm  \odot}}{\it M_{\rm *}}\Biggr)^{1/3}		\label{rtidal}
\end{equation}
(Hills 1975; Frank \& Rees 1976), the star will be totally or partially disrupted, depositing a fraction of its mass onto the BH through an accretion disc and powering a bright flare (e.g. Rees 1988; Phinney 1989; Evans \& Kochanek 1989; Lodato et al. 2009; Strubbe \& Quataert 2009; Lodato \& Rossi 2011; Guillochon \& Ramirez-Ruiz 2013, 2015a; Coughlin \& Begelman 2014). 
For a star to be disrupted outside the event horizon of a BH, that is, in order to observe the corresponding TD accretion flare, $r_{\rm t}$ must be greater than the BH event horizon radius 
\begin{equation}
r_{\rm s}=\frac{xGM_{\rm BH}}{c^2}\sim 4 \rm R_{\rm \odot} \it \Biggl(\frac{M_{\rm BH}}{\rm 10^6 \rm M_{\rm \odot}}\Biggr)\Biggl(\frac{x}{\rm 2}\Biggr),      \label{rschwar}
\end{equation}
where $x$ encapsulates effects related to the BH spin (Kesden 2010).
Hence, the non-rotating destroyer BH mass must be $M_{\rm BH}\lesssim 10^8 \rm M_{\odot}$ when solar-type stars are involved. TD accretion flares thus reveal otherwise quiescent or low-luminous BHs in a mass range complementary to that probed in AGN surveys (Vestergaard \& Osmer 2009). 

Regardless of whether the destruction is total or partial, most of the stars in a galaxy fated to be disrupted by the central BH are scattered onto low angular momentum orbits from about the BH sphere of influence, that is, onto nearly parabolic trajectories (Magorrian \& Tremaine 1999; Wang \& Merritt 2004). For this reason, in this work we assume the disruptive orbits to be parabolic.
This assumption, together with the kick naturally imparted by the disruption itself (Manukian et al. 2013), prevents our partially disrupted stars from encountering the BH a multitude of times. In this parabolic regime, about half of the stellar debris produced by a (total or partial) stellar disruption binds to the BH, returns to pericentre on different elliptical orbits (that is, with different orbital energies; Lacy et al. 1982), circularises forming an accretion disc, and falls back onto the BH emitting a peculiar flare. The fallback rate is likely to be somewhat different from the rate of debris returning to pericentre (e.g. Cannizzo et al. 1990; Ramirez-Ruiz \& Rosswog 2009; Hayasaki et al. 2013; Coughlin \& Nixon 2015; Guillochon \& Ramirez-Ruiz 2015b; Hayasaki et al. 2015; Piran et al. 2015; Shiokawa et al. 2015; Bonnerot et al. 2016; Coughlin et al. 2016), which in turn depends on the structure of the disrupted star (e.g. Lodato et al. 2009) and the properties of the encounter (e.g. Guillochon \& Ramirez-Ruiz 2013, 2015a). 

In this paper we aim at computing the critical distance $r_{\rm d}$ at which a star becomes totally or partially disrupted by the BH tidal field. We are interested in finding the critical disruption parameter 
\begin{equation}
\beta_{\rm d}=\frac{r_{\rm t}}{r_{\rm d}}=\beta \frac{r_{\rm p}}{r_{\rm d}}        \label{betad} 
\end{equation}   
for specific stellar structures that distinguishes total TDs from partial TDs, with $r_{\rm t}$ given by Eq. \ref{rtidal}, $\beta=r_{\rm t}/r_{\rm p}$ and $r_{\rm p}$ being the pericentre of the star around the BH. A partial TD is obtained for $\beta < \beta_{\rm d}$, that is, for $r_{\rm p} > r_{\rm d}$, a total TD for $\beta \geq \beta_{\rm d}$, that is, for $r_{\rm p} \leq r_{\rm d}$. The need to introduce the critical distance $r_{\rm d}$ arises because the tidal radius $r_{\rm t}$  defines where the BH tidal force overcomes the stellar self-gravity only at the stellar surface, and not everywhere within the star. 
This problem has been considered previously (Guillochon \& Ramirez-Ruiz 2013, 2015a; hereafter GRR). GRR evaluated the critical disruption parameter $\beta_{\rm d}$ for polytropes of index 5/3 and 4/3 (which represent low- and high-mass stars, respectively) using a series of adaptive mesh refinement (AMR) grid-based hydrodynamical simulations of tidal encounters of star and BH.
In this paper, we instead present the results of simulations we performed for the same purpose with the codes
\textsc{gadget2} (traditional smoothed particle hydrodynamical (SPH)\textbf{;} Springel 2005)\footnote{http://wwwmpa.mpa-garching.mpg.de/gadget/} and \textsc{gizmo} (modern SPH and mesh-free finite mass (MFM)\textbf{;} Hopkins 2015)\footnote{http://www.tapir.caltech.edu/$\sim$phopkins/Site/GIZMO.html}. Since these techniques all have advantages but also limits, we are inclined to follow GRR in finding the critical disruption parameter $\beta_{\rm d}$ for certain stellar structures\footnote{When a more realistic stellar equation of state is used (e.g. Rosswog et al. 2009, but only for white dwarfs), the value of $\beta_{\rm d}$ may change slightly.} using an MFM, a traditional SPH, and a modern SPH code instead of an AMR grid-based code, with the goal of comparing results from different techniques.

The paper is organised as follows. In Section \ref{gridVSsph} we compare AMR grid-based codes to \textsc{gizmo mfm}, traditional SPH, and modern SPH techniques. In Section \ref{loss} we discuss our method and describe how we evaluate the stellar mass loss $\Delta M$ in our simulated encounters. We show the curves of mass loss we obtain for all codes as a function of $\beta$ and polytropic index, comparing them and the corresponding $\beta_{\rm d}$ with GRR. Section \ref{conclusions} contains our summary. 

\section{Grid-based vs. SPH and \textsc{gizmo mfm} codes} \label{gridVSsph}
Fluid hydrodynamics and interactions in astrophysics are generally treated using two different classes of numerical methods: Eulerian grid-based (e.g. Laney 1998; Leveque 1998) and Lagrangian SPH (e.g. Monaghan 1992; Price 2005; Cossins 2010; Price 2012). 
Basically, grid-based methods divide a domain into stationary cells traversed over time by the investigated fluid, and account for information exchange between adjacent cells in the aim at solving the fluid equations. In particular, AMR grid-based techniques (e.g. Berger \& Oliger 1984; Berger \& Colella 1989) adapt the cell number and size according to the properties of different fluid regions, thus increasing the resolution where needed (for example in high-density regions) and reducing computational efforts and memory employment where lower resolution is sufficient. 
In contrast, SPH methods are Lagrangian by construction and model a fluid as a set of interacting fluid elements, or particles, each with its own set of fluid properties. In practice, the density of each particle is calculated by considering the neighbours within its so-called smoothing length (e.g. Price 2005), and particle velocities and entropies or internal energies are evolved according to a pressure-entropy or energy formalism (modern SPH) or a density-entropy or energy formalism (traditional SPH). Essentially, modern SPH techniques evaluate the pressure and the local density of each particle by considering the neighbours within the particle smoothing length and use pressure to define the equations of motion (Hopkins 2013). Traditional SPH techniques instead directly estimate the pressure of each particle from its local density in the same way as for the other particle properties, and use local density to define the equations of motion. In SPH methods the particle density mirrors the density of different regions of the fluid. 

Grid-based and SPH techniques both have advantages, but also limits. 
At sufficiently high velocities, grid-based methods are non-invariant under Galilean transformations, which means that different reference frames are associated with different levels of numerical diffusion among adjacent cells, and simulation results may slightly depend on the choice of the reference system (e.g. Wadsley et al. 2008). Moreover, grid-based methods violate angular momentum conservation because a fluid moving across grid cells produces artificial diffusion; this diffusion can lead to unphysical forces, which couple with the fixed structure of the grid to tie the fluid motion on specific directions (e.g. Peery \& Imlay 1988; Hahn et al. 2010). Finally, in grid-based methods hydrodynamics and gravity descriptions are mismatched, in the sense that hydrodynamics is evaluated by integrating quantities over each cell, while gravity is computed at the centre of each cell and then interpolated at the desired position (as for collisionless particles). This can produce spurious instabilities (e.g. Truelove et al. 1997). 

SPH methods first need an artificial viscosity term added to the particle equation of motion in order to resolve shocks (Balsara 1989; Cullen \& Dehnen 2010). Second, traditional SPH codes are associated with a surface tension between fluid regions of highly different densities, which limits their mixing (e.g. Agertz et al. 2007).
Great effort has been made to improve SPH methods, leading to the so-called modern SPHs (Hopkins 2013). The smoothed definition of pressures together with densities, the more sophisticated viscosity switches, the higher order smoothing kernels (quintic spline instead of cubic spline; see below), and the inclusion of artificial conduction allowed solving these problems, at least partially. However, the higher order kernels typically lead to excessive diffusion. Despite all these improvements, some intrinsic limits of this technique still remain, such as the ideal infinite number of neighbours required to capture small-amplitude instabilities.

\begin{figure*}
\centering
\vspace{0.2cm}
\includegraphics[width=19.cm, angle=0]{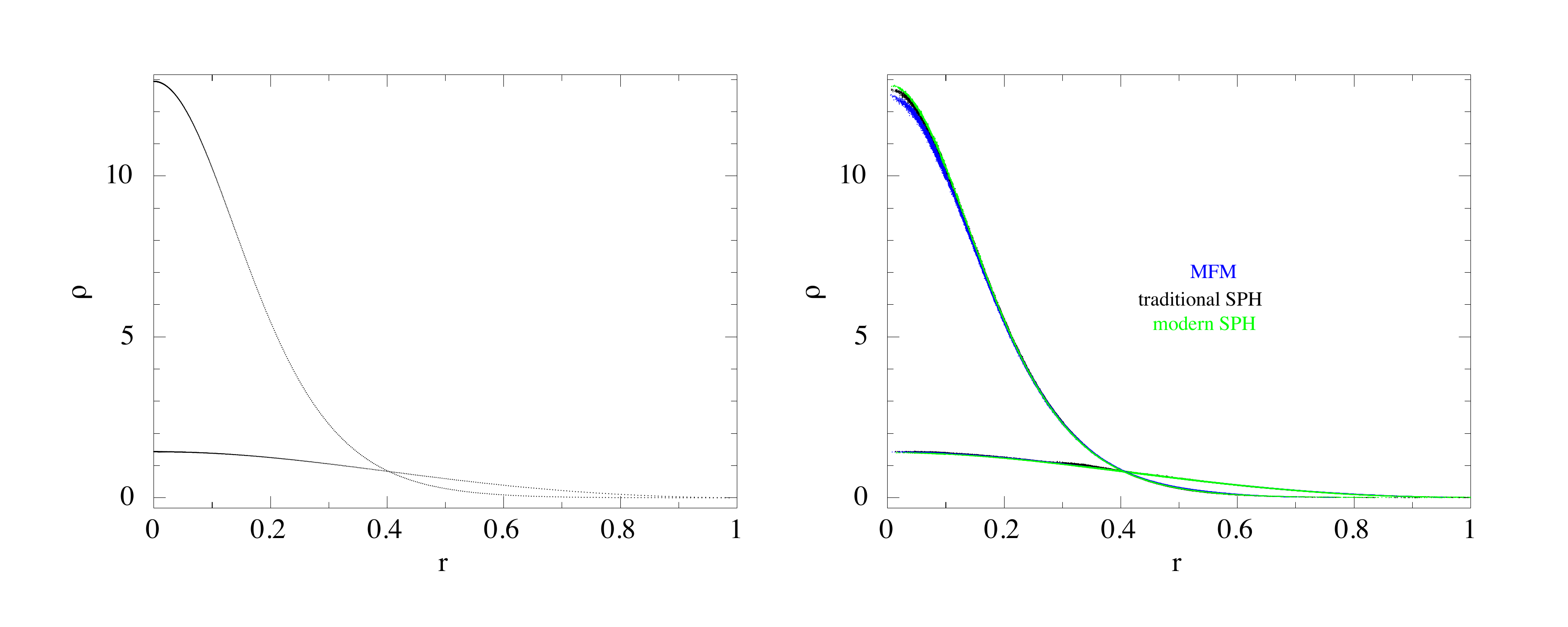} \\
\caption{Left panel: Analytic solutions for the $\gamma=4/3, 5/3$ polytropic radial density profile from the Lane-Emden equation. Right panel: Plot of the relaxed radial stellar density profile for each simulation technique for both politropic indices ($\gamma=4/3, 5/3$ from the highest to the lowest central density). Units are $\rm M_{\rm \odot}/\rm R_{\rm \odot}^3$ for $\rho$ and $\rm R_{\rm \odot}$ for $r$.  \label{densities}}
\end{figure*}
\begin{figure*}
\centering
\vspace{0.2cm}
\includegraphics[width=11.cm, angle=0]{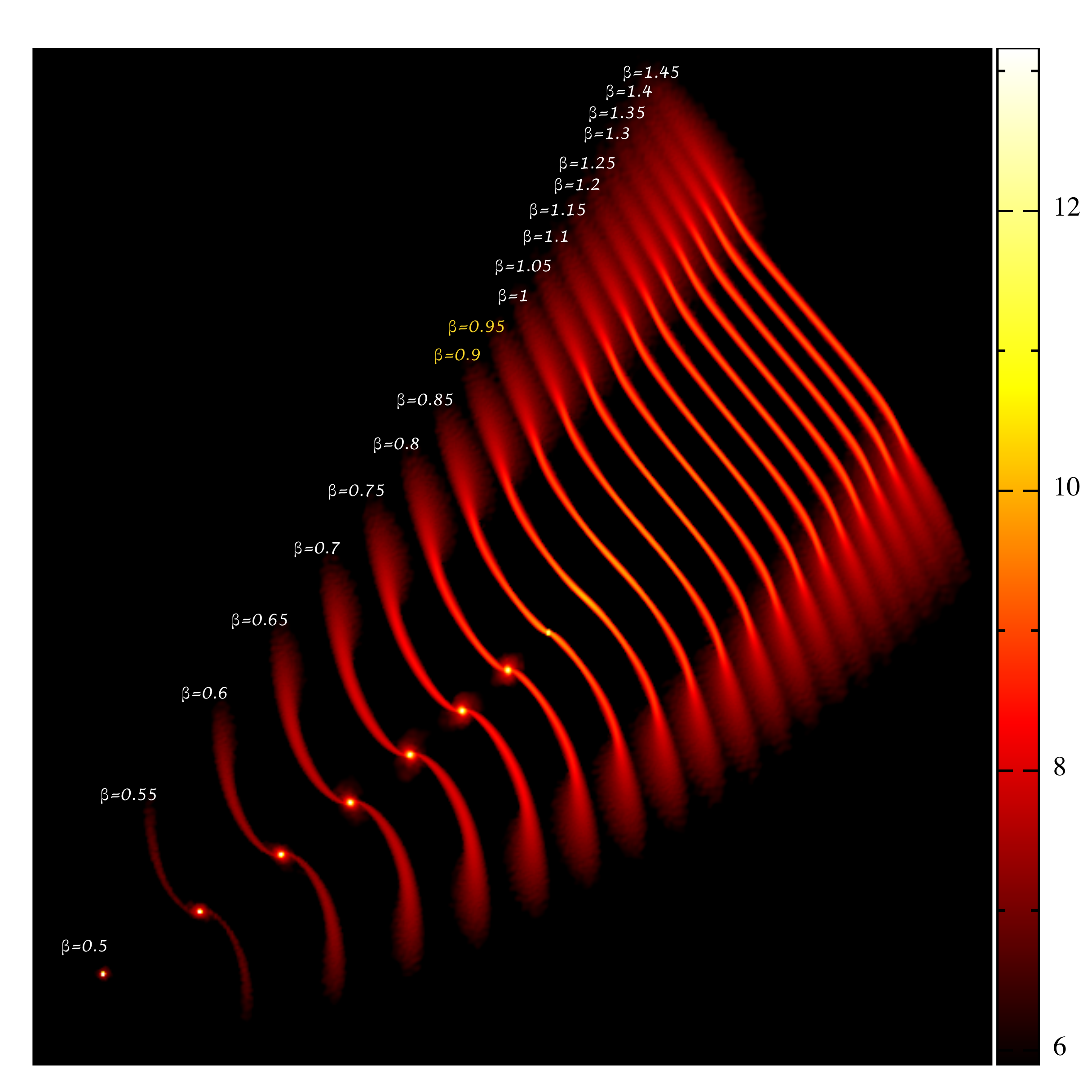} \\
\caption{Snapshots of the SPH particle density (in logarithmic scale) at $t\sim 8.5\times 10^4\rm s$ after pericentre passage for our \textsc{gadget2} simulations, in the case of a star with polytropic index $5/3$. White and black correspond to the highest and lowest densities, respectively. Each snapshot is labelled with the corresponding value of $\beta$, with the range where the critical disruption parameter $\beta_{\rm d}$ lies highlighted in yellow. \label{53_stars}}
\end{figure*}
\begin{figure*}
\centering
\vspace{0.2cm}
\includegraphics[width=11.cm, angle=0]{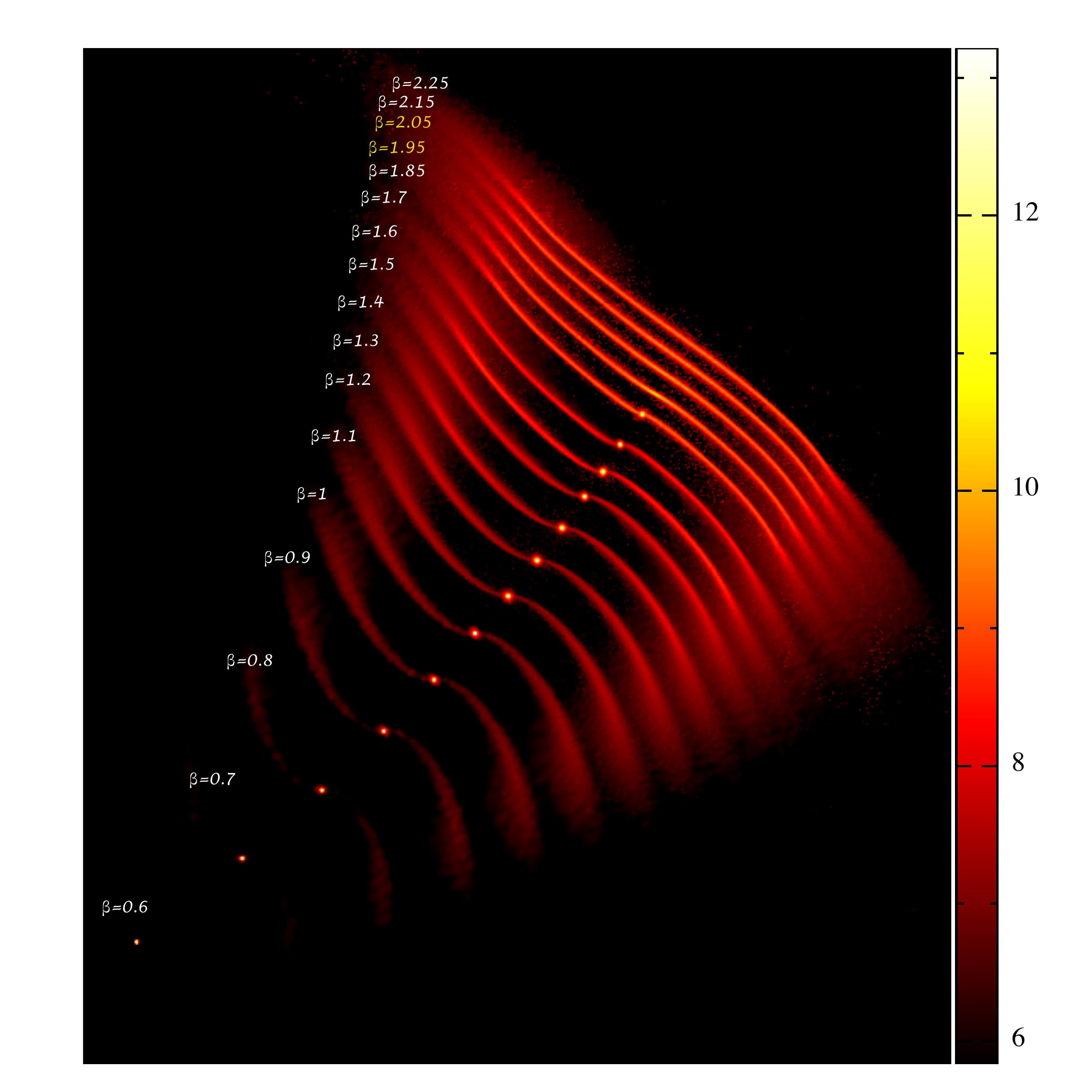} \\
\caption{Same as Fig. \ref{53_stars} for a polytropic star of index $4/3$.  \label{43_stars}}
\end{figure*}

Recently, a completely new Lagrangian method that aims to simultaneously capture the advantages of both SPH and grid-based techniques, has been implemented in the public code \textsc{gizmo} (Hopkins 2015). In \textsc{gizmo}, the volume is discretised among a discrete set of tracers (particles) through a partition scheme based on a smoothing kernel (in a way that is similar to SPH codes). However, unlike SPH codes, these particles do not sample fluid elements, but only represent the centre of unstructured cells that are free to move with the fluid, like in moving mesh codes (Springel 2010). Hydrodynamics equations are then solved at the cell boundaries, defined by an effective face. This guarantees an exact conservation of energy and linear and angular momentum as well as an accurate description of shocks without needing an artificial viscosity term. The density associated with each particle or cell is obtained by dividing the mass of the cell for its effective volume. In this work, we use the mesh-free finite mass method of \textsc{gizmo}, where particle mass is preserved, making the code perfectly Lagrangian. For this method,  we use the cubic spline kernel with a desired number of neighbours equal to 32 for the partition.

\section{SPH and \textsc{gizmo mfm} simulations and stellar mass losses} \label{loss}
We modelled stars as polytropes of index 5/3 (low-mass stars) or 4/3 (high-mass stars), with masses and radii of  $1 \rm M_{\rm \odot}$ and $1 \rm R_{\rm \odot}$, sampling each of them with $N_{\rm part}\sim10^5$ particles. This is done by placing the particles through a close sphere packing and then stretching their radial positions to reach the required polytropic density profile, thus limiting the statistical noise associated with a random placement of the particles.  
$N_{\rm part}$ sets the gravitational softening length of each particle in our codes to $\epsilon \sim 0.1R_{\rm *}/(N_{\rm part})^{1/3}\sim 0.002 \rm R_{\rm \odot}$, preventing particle overlapping in evaluating gravitational interactions. We also tried test runs at higher resolution, where we modelled stars with $\sim10^6$ particles, but did not find significant differences in the stellar mass loss $\Delta M$ with respect to simulations with lower resolution.
We evolved stars in isolation for several dynamical times in order to ensure their stability. The right panel of Fig. \ref{densities} shows the relaxed stellar density profile, that is, the local density of the particles $\rho (r)$ (in $\rm M_{\rm \odot}/\rm R_{\rm \odot}^3$) versus their radial distance from the stellar centre of mass $r$ (in $\rm R_{\rm \odot})$, for each simulation technique for the two polytropic indices ($\gamma=4/3$, and $5/3$ from the highest to the lowest central density), compared to the analytic solutions from the Lane-Emden equation (left panel). The kernel function that drives the evaluation of each particle local density (e.g. Price 2005) and the volume partition (Hopkins 2015) is chosen to be a cubic (in \textsc{gadget2} and \textsc{gizmo mfm}) or quintic (in \textsc{gizmo} modern SPH) spline, and the number of neighbours of each particle and domain point within its smoothing length/kernel size is fixed to 32 and 128, respectively (Monaghan \& Lattanzio 1985; Hongbin \& Xin 2005; Dehnen \& Aly 2012). Gravitational forces are computed through the Springel relative criterion (Springel 2005) instead of the standard Barnes-Hut criterion (Barnes \& Hut 1986) because the Springel criterion shows better accuracy at the same computational cost. Since the relative criterion is based on the particle acceleration, which is not available at the beginning of each simulation, the Barnes-Hut criterion is adopted at the first timestep to estimate an acceleration value, and then the iteration is repeated using the Springel criterion in order to remain consistent with the subsequent iterations. In our simulations we use quite a large opening angle value (0.7), but the accuracy parameter for the relative criterion is set to 0.0005, which is very small compared to the suggested standard value (0.0025). We performed test runs setting the opening angle to 0.1 and increasing the accuracy parameter to 0.0025, but found no differences in the stellar density and temperature profiles and in $\Delta M$.
We implemented the BH force through a Newtonian analytical potential, with $M_{\rm BH}=10^6 \rm M_{\rm \odot}$, and in each of the traditional SPH, modern SPH and \textsc{gizmo mfm} simulations we placed one star on a parabolic orbit with a given $r_{\rm p}$, that is, $\beta$, around the BH. The star was initially placed at a distance five times greater than $r_{\rm t}$ to avoid spurious tidal distortions (we also tested larger initial distances, but found no significant differences in the outcomes). Stellar rotation is not expected to significantly affect our results in the range of $\beta$ considered in this paper (Stone et al. 2013). Figures \ref{53_stars} and \ref{43_stars} show snapshots from our traditional SPH simulations recorded shortly after pericentre passage. The lower limit of the range where $\beta_{\rm d}$ lies (yellow) allows the core recollapse to occur for both polytropic indices (Guillochon \& Ramirez-Ruiz 2013, 2015a).  
Modern SPH and \textsc{gizmo mfm} simulations give almost the same results. 
\begin{figure*}
\centering
\includegraphics[width=13.cm, angle=0]{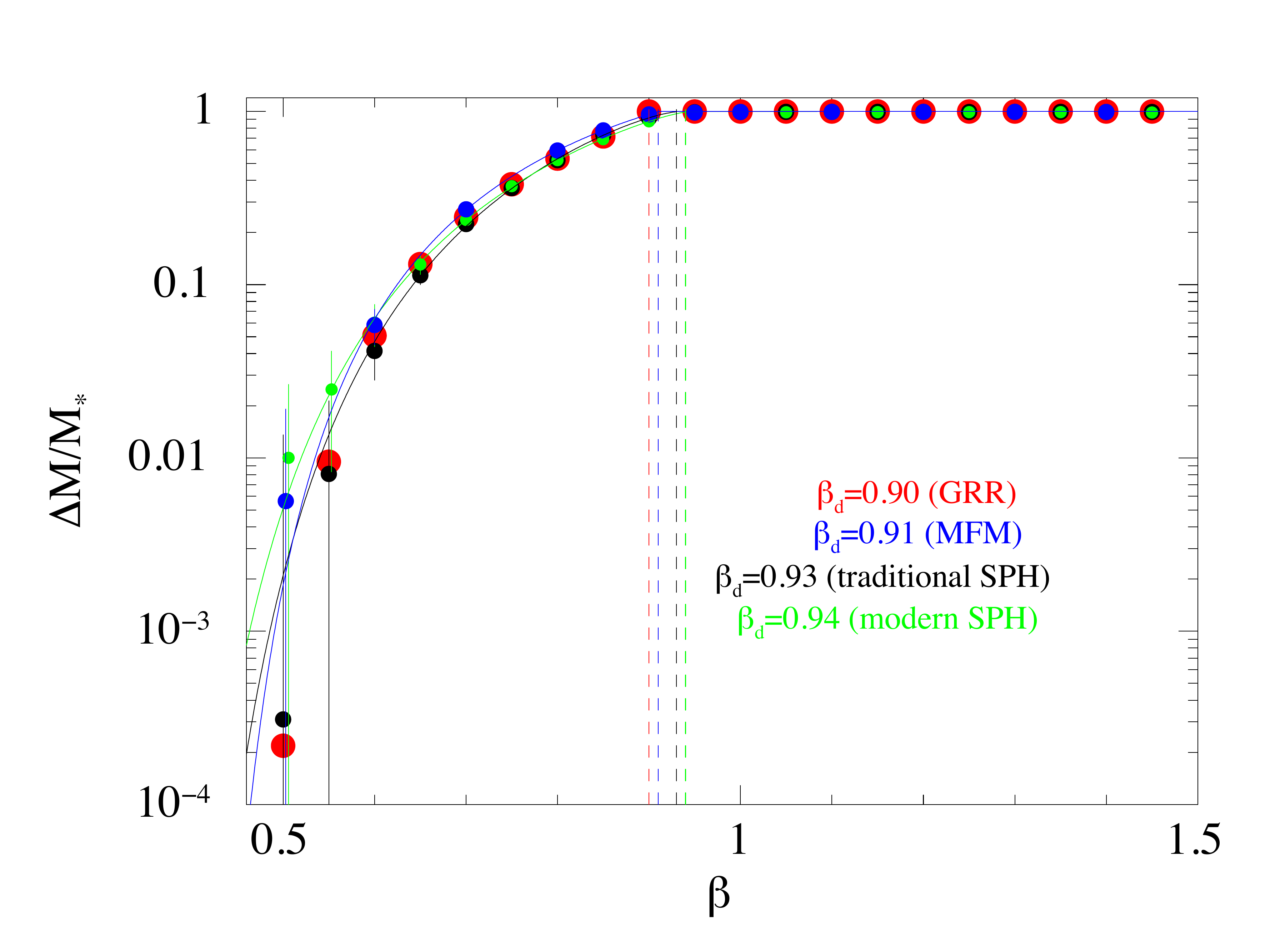} \\
\caption{Stellar mass loss (in units of $\Delta M/M_{\rm *}$) as a function of $\beta$ for a star with polytropic index 5/3. $\Delta M$ is evaluated at $t\sim 10^6 \rm s$ after the disruption. Blue, black, green, and red points are associated with \textsc{gizmo mfm}, \textsc{gadget2}, \textsc{gizmo} modern SPH, and GRR simulations, respectively. Uncertainties on $\Delta M/M_{\rm *}$ from SPH and \textsc{gizmo mfm} simulations are inferred as reported in the main text. Points at low values of $\beta$ have been slightly horizontally displaced to give a better view of the error bars. The value of the critical disruption parameter $\beta_{\rm d}$ (dashed lines) slightly depends on the adopted simulation method.  \label{53_curve}}
\end{figure*}
\begin{figure*}
\centering
\includegraphics[width=13.05cm, angle=0]{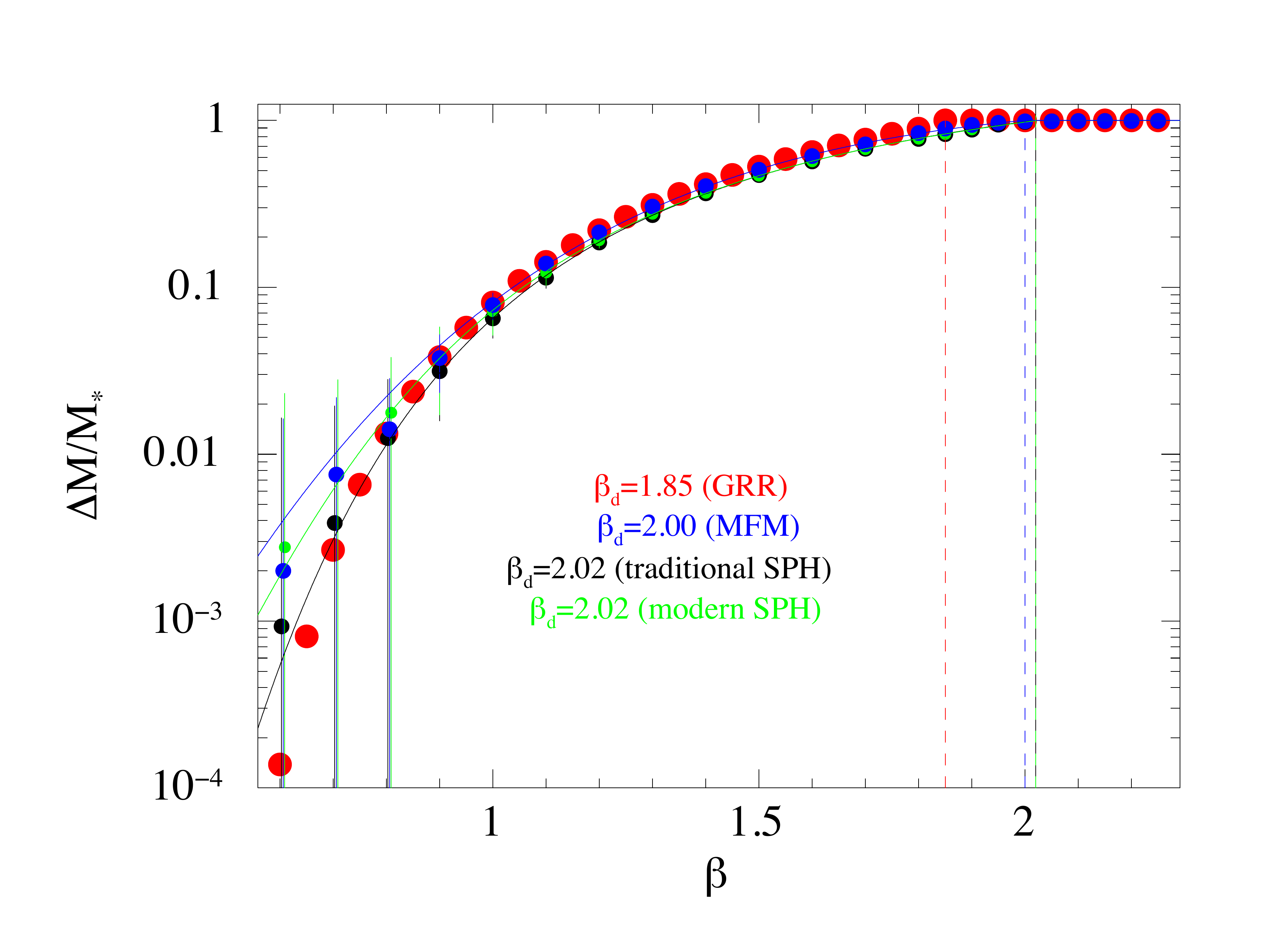} \\
\caption{Same as Fig. \ref{53_curve} for a polytropic star of index 4/3. The values of $\beta_{\rm d}$ obtained from our simulations visibly differ from those of GRR. \label{43_curve}}
\end{figure*}

We aim to assess the stellar mass loss $\Delta M$ in each simulation. We recall that $\Delta M=M_{\rm *}$ corresponds to total disruption.
\begin{table*}
\centering
\caption{Fitting coefficients of Eq. \ref{ff1} and $\beta_{\rm d}$ for each of the point sets in Figs. \ref{53_curve} and \ref{43_curve}. \label{fitting_table}}
\begin{tabular}{c|c|c|c|c|c|c|c}
Simulations & Polytropic index & \it A & \it B & \it C & \it D & \it E & $\beta_{\rm d}$ \\
& & & & & & & \\
\hline
& & & & & & & \\
GRR & 5/3 & 3.1647 & -6.3777 & 3.1797 & -3.4137 & 2.4616 & 0.90 \\
\textsc{gizmo mfm} & 5/3 & 5.4722 & -11.764 & 6.3204 & -3.8172 & 2.8919 & 0.91 \\
\textsc{gadget2} & 5/3 & 8.9696 & -19.111 & 10.180 & -4.2964 & 3.3231 & 0.93 \\
\textsc{gizmo} modern SPH & 5/3 & 8.7074 & -18.358 & 9.6760 &-4.5340 & 3.5914 & 0.94 \\
GRR & 4/3 & 12.996 & -31.149 & 12.865 & -5.3232 & 6.4262 & 1.85 \\
\textsc{gizmo mfm} & 4/3 & -13.964 & 11.217 & -2.1168 & 0.3930 & 0.5475 & 2.00\\
\textsc{gadget2} & 4/3 & -15.378 &-5.2385 & 6.3635 & -1.5122 & 5.7378 & 2.02 \\
\textsc{gizmo} modern SPH & 4/3 & -10.394 & -0.2160 & 2.6421 & -0.8804 & 2.9215 & 2.02 \\
\end{tabular}
\end{table*}
We describe the method we adopted to evaluate $\Delta M$ from each of our simulated star-BH tidal encounters, following GRR. 
In a specific simulation at a specific time, the position and velocity components of the stellar centre of mass around the BH, $x_{\rm CM}$, $y_{\rm CM}$, $z_{\rm CM}$, $v_{\rm x_{\rm CM}}$, $v_{\rm y_{\rm CM}}$, and $v_{\rm z_{\rm CM}}$ are defined through an iterative approach. As a first step, we choose them to coincide with the position and velocity components of the particle with the highest local density, $x_{\rm peak}$, $y_{\rm peak}$, $z_{\rm peak}$, $v_{\rm x_{\rm peak}}$, $v_{\rm y_{\rm peak}}$, $v_{\rm z_{\rm peak}}$. The specific binding energy to the star of the $i$-$\rm th$ particle then reads
\begin{equation}
E_{\rm *_{\it i}}=\frac{1}{2}\Bigl[(v_{\rm x_{\it i}}-v_{\rm x_{\rm peak}})^2+(v_{\rm y_{\it i}}-v_{\rm y_{\rm peak}})^2+(v_{\rm z_{\it i}}-v_{\rm z_{\rm peak}})^2\Bigr]+\phi_{\rm *_{\it i}},          \label{Estar}
\end{equation}
where $v_{\rm x_{\it i}}$, $v_{\rm y_{\it i}}$, and $v_{\rm z_{\it i}}$ are the velocity components of the $i$-$\rm th$ particle and $\phi_{\rm *_{\it i}}$ the stellar gravitational potential acting on the $i$-$\rm th$ particle (directly computed by the simulation code). By considering only particles with $E_{\rm *_{\it i}}< 0$, we re-define the position and velocity components of the star centre of mass and re-evaluate Eq. \ref{Estar} by setting them in place of the components labelled with the subscript "peak". The process is re-iterated until the convergency of $v_{\rm CM}$ to a constant value to lower than $10^{-5} \rm R_{\rm \odot} \rm yr^{-1}$. Particles with $E_{\rm *_{\it i}}> 0$ are unbound from the star. The stellar mass loss at the considered time can be obtained by multiplying the mass of a single particle, $m=M_{\rm *}/\it N_{\rm part}$, by the number of particles bound to the star, $N_{\rm Bound}$, and subtracting the result from $M_{\rm *}$. $\Delta M$ is obtained at $t\sim 10^6 \rm s$ ($\sim650$ stellar dynamical times) after the disruption. 
\begin{table}
\centering
\caption{$\beta_{\rm d}$ value as a function of polytropic index and adopted simulation method. \label{beta_table}}
\begin{tabular}{c|c|c}
Simulation method & Polytropic index & $\beta_{\rm d}$ \\
& &  \\
\hline
& & \\
AMR grid-based & 5/3 & 0.90 \\
MFM & 5/3 & 0.91 \\
Traditional SPH & 5/3 & 0.93 \\
Modern SPH & 5/3 & 0.94 \\
AMR grid-based & 4/3 & 1.85 \\
MFM & 4/3 & 2.00 \\
Traditional SPH & 4/3 & 2.02 \\
Modern SPH & 4/3 & 2.02 \\
\end{tabular}
\end{table}

Figures \ref{53_curve} and \ref{43_curve} show the stellar mass loss in units of $\Delta M/M_{\rm *}$ as a function of $\beta$  for polytropes of index $5/3$ and $4/3$, respectively, inferred from our simulations with \textsc{gizmo mfm} (blue points), \textsc{gadget2} (black points) and \textsc{gizmo} modern SPH (green points), and the same obtained from the GRR simulations (red points). We estimate the uncertainty on our inferred $\Delta M/M_{\rm *}$ as
\begin{equation}
\sigma_{\frac{\rm \Delta M}{M_{\rm *}}}=\sqrt{\sigma_{\rm Poisson}^2+\sigma_{\rm E_{\rm *_{\rm i}}}^2+\sigma_{\rm AD}^2}=\sqrt{\Biggl(\frac{\sqrt{N_{\rm Bound}}}{N_{\rm part}}\Biggr)^2+0.01^2+\sigma_{\rm AD}^2}
\end{equation}
where $\sigma_{\rm AD}$ is the average deviation from 1 of $\Delta M/M_{\rm *}$ for total disruptions in each of our point sets and $\sigma_{\rm E_{\rm *_{\rm i}}}=0.01$, as the values of $|E_{\rm *_{\rm i}}|$ for about $10^3$ particles of $10^5$ are lower than 0.01 times the average value $\overline{|E_{\rm *}|}$, that is, we are not able to determine exactly whether these $10^3$ particles are bound to or unbound from the star. 
We fit each of our point sets with a function introduced in GRR
\begin{equation*}
f(\beta)=\exp{\Biggl[\frac{A+B\beta+C\beta^2}{1-D\beta+E\beta^2}\Biggr]} , \; \; \; \; \; \beta<\beta_{\rm d}  \label{ff}
\end{equation*}  
\begin{equation}
f(\beta)=1, \; \; \; \; \; \beta \geq \beta_{\rm d}. \label{ff1}
\end{equation}
The values of the coefficients $A, B, C, D$, and $E$ and of $\beta_{\rm d}$ are given in Table \ref{fitting_table}. 
 \begin{figure*}
\centering
\includegraphics[width=18.5cm, angle=0]{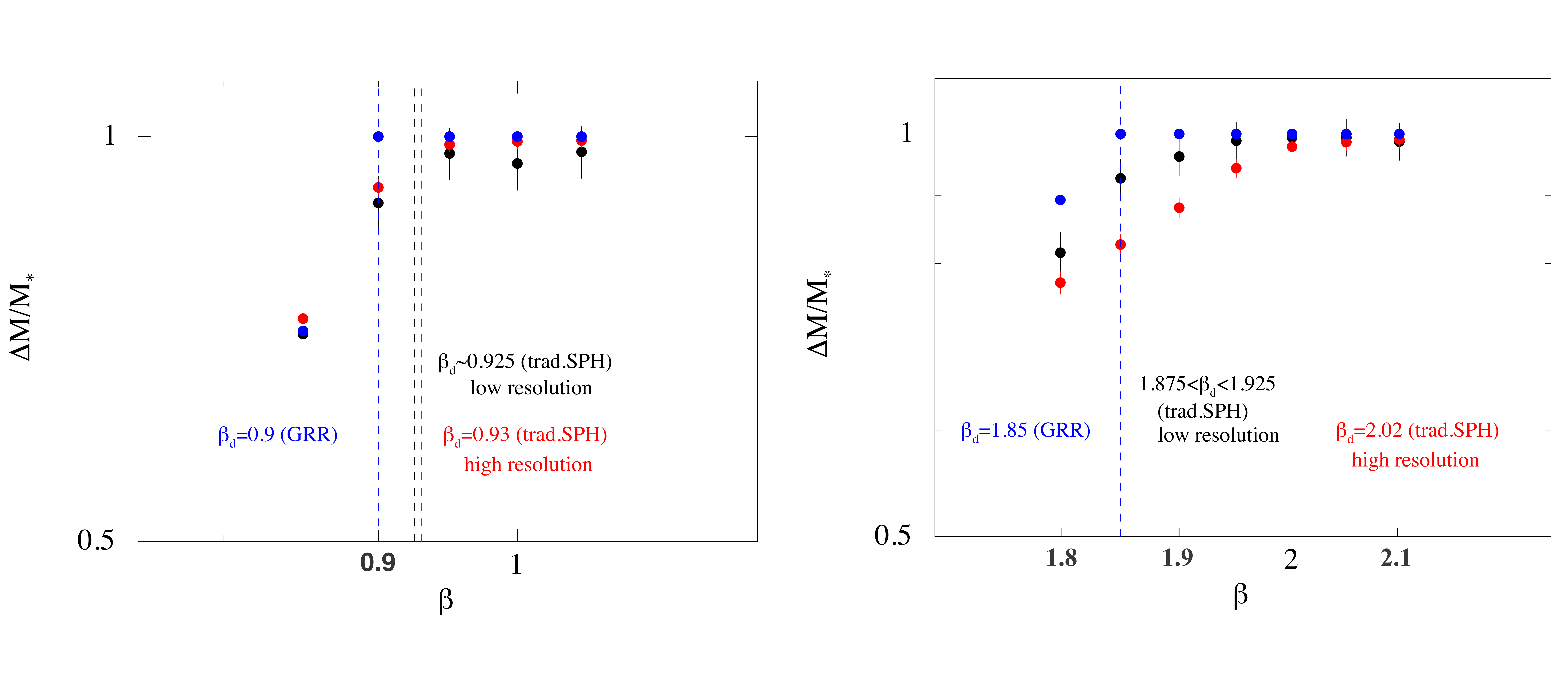} \\
\caption{Comparison of mass losses as a function of $\beta$ near $\beta_{\rm d}$ between the GRR simulations (blue points), high- ($\sim 10^5$ particles; red points) and low-resolution ($\sim 10^3$ particles; black points) \textsc{gadget2} simulations, for $\gamma=5/3$ (left panel) and $\gamma=4/3$ (right panel) polytropes. For a $\gamma=4/3$ polytrope, the value of $\beta_{\rm d}$ clearly depends on the adopted resolution below a resolution threshold. For a $\gamma=5/3$ polytrope, the value of $\beta_{\rm d}$ differs very slightly among the three simulations. \label{res_com}}
\end{figure*}
It is worth noting that for the 5/3 polytropic index the curves of stellar mass loss associated with the four simulation codes differ very slightly in the value of the critical disruption parameter $\beta_{\rm d}$ (dashed lines in Fig. \ref{53_curve}). Specifically, $\beta_{\rm d}$ is reached first in the GRR simulations ($\beta_{\rm d}$=0.90), followed by the \textsc{gizmo mfm} ($\beta_{\rm d}=$0.91), \textsc{gadget2} ($\beta_{\rm d}=$0.93), and \textsc{gizmo} modern SPH ($\beta_{\rm d}=$0.94) simulations (Table \ref{beta_table}). This is expected because of the greater degree of excessive diffusion that characterises grid-based techniques compared to modern and traditional SPH techniques and the surface tension conversely involved in SPH methods (Section \ref{gridVSsph}). For the 4/3 polytropic index, instead, there is disagreement between our simulations and those of GRR (dashed lines in Fig. \ref{43_curve}). $\beta_{\rm d}$ is reached clearly first in the simulations of GRR ($\beta_{\rm d}$=1.85), followed by very similar values of the \textsc{gizmo mfm} ($\beta_{\rm d}=$2.00), \textsc{gadget2} ($\beta_{\rm d}=$2.02), and \textsc{gizmo} modern SPH ($\beta_{\rm d}=$2.02) simulations (Table \ref{beta_table}).
We hypothesise that the lower value of $\beta_{\rm d}$ obtained by GRR is the result of resolving the stellar core of the $\gamma=4/3$ polytrope not far enough.

In support of this hypothesis, we tested the dependence of $\beta_{\rm d}$ on the resolution of our simulations by performing some low-resolution ($\sim 10^3$ particles) \textsc{gadget2} simulations for the two polytropic indices (black points in Fig. \ref{res_com}). Fig. \ref{res_com} shows that for a $\gamma=5/3$ polytrope (left-hand panel) the change in resolution has negligible effects on $\beta_{\rm d}$. On the other hand, for $\gamma=4/3$ polytropes (right-hand panel) we observe a strong dependence of $\beta_{\rm d}$ on resolution below a resolution threshold because the configuration of the star is less stable.

\begin{figure}[h!]
\centering
\includegraphics[width=9.cm, angle=0]{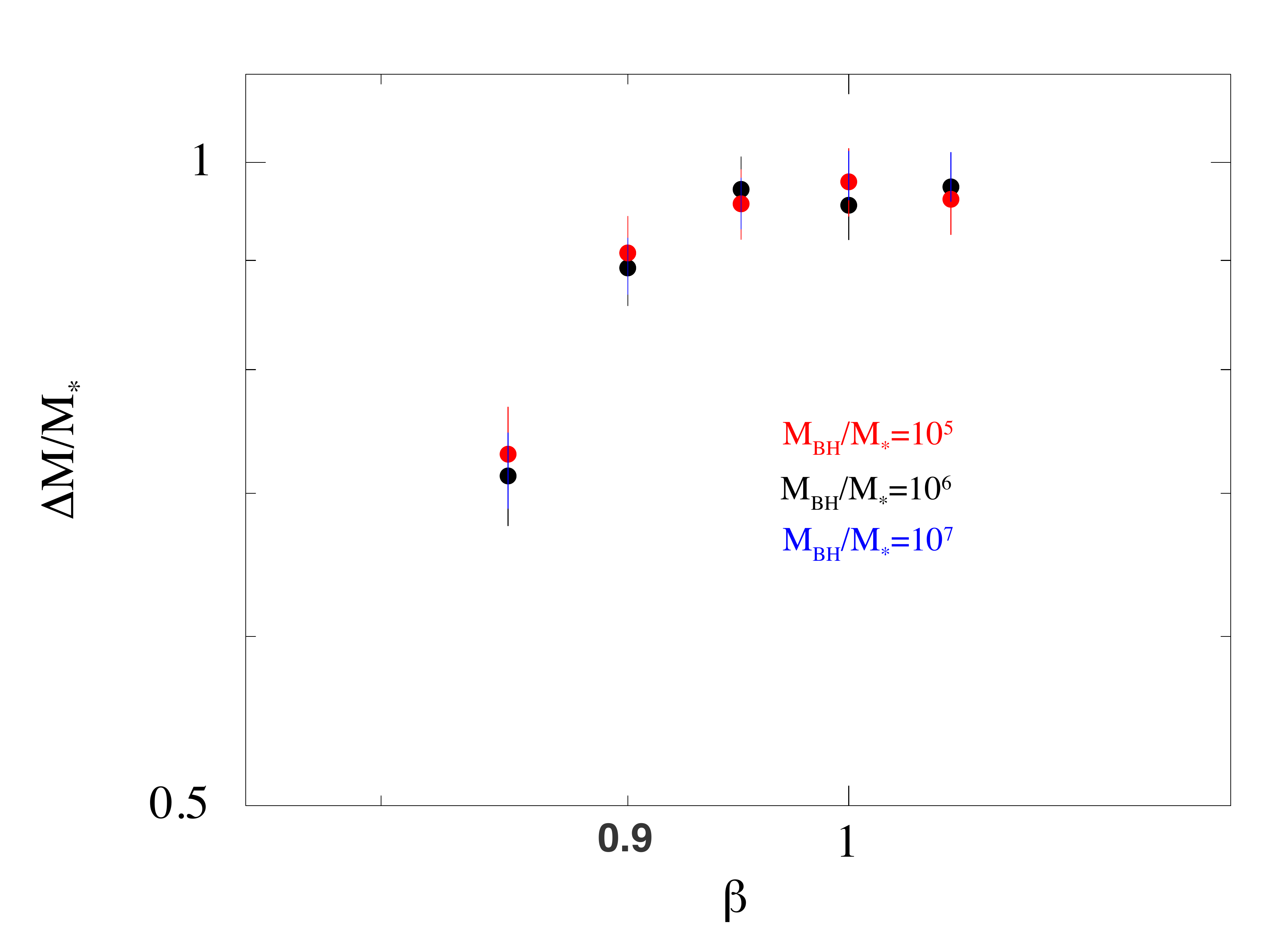} \\
\caption{Comparison of mass losses as a function of $\beta$ near $\beta_{\rm d}$ for a $\gamma=5/3$ polytrope of mass $1 \rm M_{\rm \odot}$ approaching BHs with three different masses: $10^5 \rm M_{\rm \odot}$ (red points), $10^6 \rm M_{\rm \odot}$ (black points), $10^7 \rm M_{\rm \odot}$ (blue points). The value of $\beta_{\rm d}$ clearly does not depend on $M_{\rm BH}$. \label{MBH_com}}
\end{figure}

 We also determined the dependence of $\beta_{\rm d}$ on different values of $M_{\rm BH}$ by performing additional low-resolution ($\sim 10^3$ particles) \textsc{gadget2} simulations with a $\gamma=5/3$ polytrope of mass $1 \rm M_{\rm \odot}$ and BHs of masses $10^5 \rm M_{\rm \odot}$ and $10^7 \rm M_{\rm \odot}$. Fig. \ref{MBH_com} clearly shows that $\beta_{\rm d}$ does not depend sensitively on $M_{\rm BH}$. We recall that flares and accretion temperatures instead depend on $M_{\rm BH}$ (e.g. Guillochon \& Ramirez-Ruiz 2013, 2015a).

\begin{figure*}
\centering
\includegraphics[width=18.cm, angle=0]{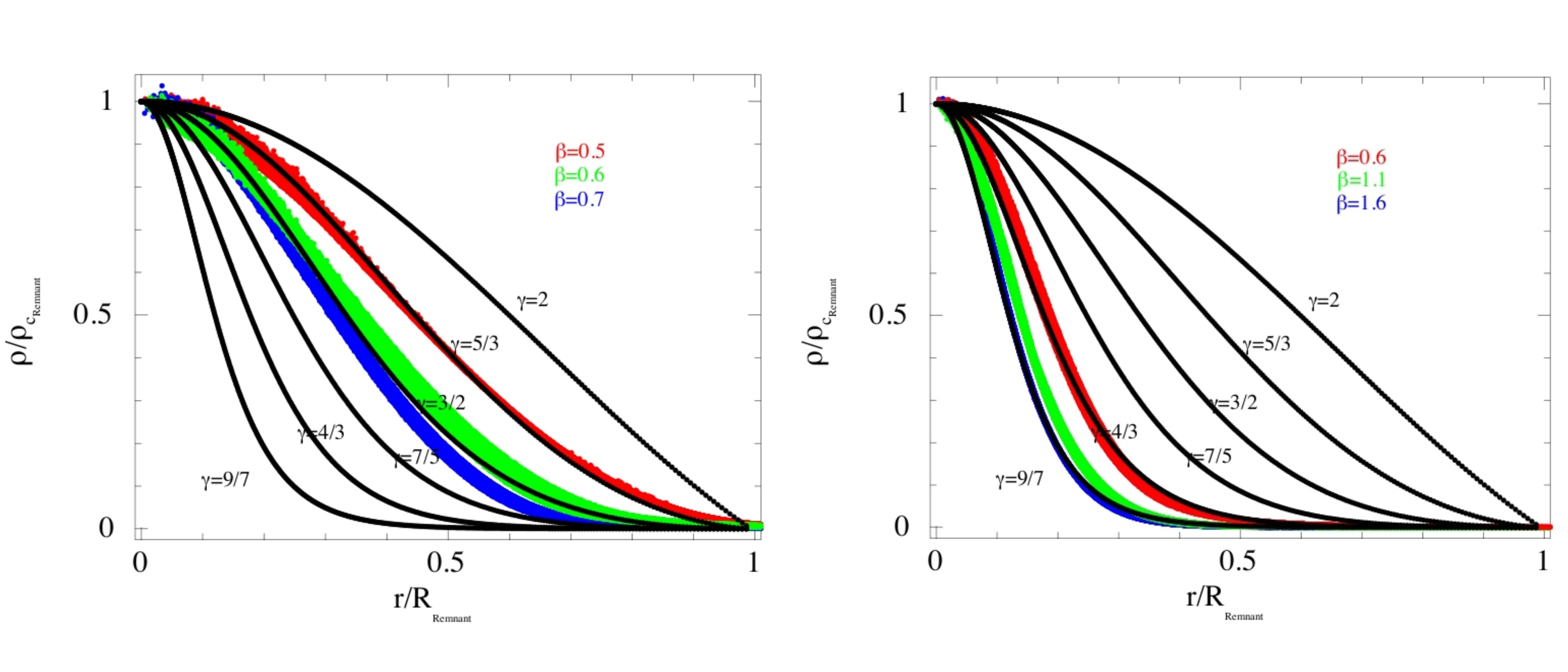} \\
\caption{Changes in the value of the polytropic index of the stellar remnant resulting from partial disruptions for selected initial values of its $\beta$. Densities and radii are normalised to the central density and the radius of the remnant. Black curves represent solutions to the Lane-Emden equation for different values of $\gamma$; red, green, and blue points are from some of our simulations that left a remnant, for three different values of $\beta$. Left panel: $\gamma=5/3$ polytrope. Right panel: $\gamma=4/3$ polytrope. \label{n_com}}
\end{figure*}

For completeness, we also show in Fig. \ref{n_com} how the polytropic index of the stellar remnant, which results from partial disruptions on parabolic orbits, is not preserved, but decreases with increasing $\beta$ for both $\gamma=5/3$ polytropes (left panel) and $\gamma=4/3$ polytropes (right panel).

\section{Summary and conclusions} \label{conclusions}
Tidal disruption events provide a unique way to probe otherwise quiescent or low-luminous black holes at the centres of galaxies. When approaching the central black hole of a galaxy, a star may be totally or partially disrupted by the black hole tidal field, depositing material onto the compact object and lighting it up through a bright flare (e.g. Rees 1988; Phinney 1989; Evans \& Kochanek 1989). Such a tidal accretion flare is expected to be shaped by the structure of the disrupted star (e.g. Lodato et al. 2009) and the morphology of the star-black hole encounter (e.g. Guillochon \& Ramirez-Ruiz 2013, 2015a). 

The hydrodynamical simulations of Guillochon \& Ramirez-Ruiz of star-black hole close encounters probably represent the most complete theoretical investigation of the properties of tidal disruption events (Guillochon \& Ramirez-Ruiz 2013, 2015a). In each simulation, the star ($M_{\rm *}=1\rm M_{\rm \odot}$, $R_{\rm *}=1\rm R_{\rm \odot}$) is modelled as a polytrope of index 5/3 or 4/3 and evolved on a parabolic orbit with a specific pericentre around the black hole ($M_{\rm BH}=10^6 \rm M_{\rm \odot}$) using an AMR grid-based code. The resulting stellar mass loss defines the morphology of the simulated encounter, that is, it defines whether the disruption is total or partial, thus shaping the ensuing accretion flare.  
Here we followed the approach of Guillochon \& Ramirez-Ruiz, but adopted two SPH simulation codes (\textsc{gadget2}, traditional SPH; Springel 2005; \textsc{gizmo}, modern SPH; Hopkins 2015) and \textsc{gizmo} in \textsc{mfm} mode (Hopkins 2015) instead of a grid-based method, as all these simulation techniques have their advantages, but also limits (Section \ref{gridVSsph}). 
We mainly intended to determine for each polytropic index whether the demarcation line between total and partial tidal disruption events, the critical disruption parameter $\beta_{\rm d}$ (Eq. \ref{betad}), is the same for different simulation techniques. 

Figs. \ref{53_curve} and \ref{43_curve} clearly show that for a $\gamma=5/3$ polytrope the curves of stellar mass loss inferred from AMR grid-based simulations (red points) and from \textsc{gizmo mfm} (blue points), traditional SPH (black points), and modern SPH (green points) simulations differ only slightly in the value of $\beta_{\rm d}$ (dashed lines), reflecting the limits of different codes (Section \ref{gridVSsph}), while for a $\gamma=4/3$ polytrope there is disagreement between our simulations and those of GRR (Table \ref{beta_table}), which is most likely due to the adopted resolutions; this interpretation is consistent with the resolution tests we performed with our own simulations (Fig. \ref{res_com}). However, even with equal resolution, the SPH approach should be superior to a grid-based approach at resolving the dynamics of the core of, especially, a $\gamma=4/3$ polytrope, given that the resolution naturally follows density in equal-mass-particle approaches. As a consequence, we find $\beta_{\rm d}=0.92\pm 0.02$ ($2.01\pm 0.01$) for a $\gamma=5/3$ ($4/3$) polytrope. 

The $\gamma=4/3$ profile is probably only appropriate for a zero-age main-sequence sun because the central density of our Sun is about twice greater than the $\gamma=4/3$ polytrope at an age of 5 Gyr. For a real star, even greater resolution would therefore be needed in a grid-based approach in order to properly estimate the location of full versus partial disruption. Moreover, real stars are generally not well modelled by a single polytropic index, especially as they evolve (MacLeod et al. 2012). Giant stars consist of a tenuous envelope and a dense core, which prevents envelope mass loss, thus likely moving the value of $\beta_{\rm d}$ even ahead. A similar core-envelope structure and behaviour also characterise giant planets when they are disrupted by their host star (Liu et al. 2013). TDEs could also refer to disruptions by stellar objects (Guillochon et al. 2011; Perets et al. 2016). However, the value of $\beta_{\rm d}$ for the latter encounters still remains to be investigated.

\begin{acknowledgements}
We thank the ISCRA staff for allowing us to perform our simulations on the Cineca Supercomputing Cluster GALILEO. 
We acknowledge P. F. Hopkins and G. Lodato for very useful discussion and comments on this work. 
ERC acknowledges support by NASA through the Einstein Fellowship Program, grant PF6-170150.
This work was also supported by the Packard grant and NASA ATP grant NNX14AH37G (ER).
We also thank the referee, H. B. Perets, for valuable comments on the manuscript and very constructive suggestions.
\end{acknowledgements}

\end{document}